\PassOptionsToPackage{capitalise}{cleveref}
\PassOptionsToPackage{svgnames,dvipsnames}{xcolor}
\documentclass[runningheads]{llncs}

\usepackage{hyperref}
\usepackage[T1]{fontenc}
\usepackage{fancyvrb}
\usepackage{fvextra}
\usepackage{subcaption}
\usepackage{cite}

\usepackage{tikz}

\usepackage{mathtools}

\usepackage{bm}

\usepackage{amsfonts}

\usepackage{amssymb}

\usepackage{wrapfig}

\usepackage{extpfeil}

\usepackage{etoolbox}

\usepackage{xcolor}

\usepackage{booktabs}

\usepackage{cases}

\usepackage{semantic}

\usepackage{relsize}

\usepackage{pgf}

\usepackage{cleveref}

\usepackage{sessiontypes} \newcommand{\Eg}{\textit{E.g.,}\xspace}

\newcommand{\etal}{\textit{et al.}\xspace}
\newcommand{\sut}{SUT\xspace}
\newcommand{\tool}{\textsc{COTS}\xspace}
\newcommand{\model}{\textsc{COpenAPI}\xspace}

\newcommand{\equal}{\mathrel{=}}

\newcommand{\lsem}{[\![}
\newcommand{\rsem}{]\!]}

\usepackage{xstring}

\usepackage{xspace}

\usepackage{pifont}

\usepackage{nicefrac}

\usepackage[inline,shortlabels]{enumitem}

\definecolor{yellowhighlight}{RGB}{255, 255, 64}
\definecolor{orangehighlight}{RGB}{255, 204, 153}

\definecolor{dkgreen}{rgb}{0,0.6,0}
\definecolor{gray}{rgb}{0.5,0.5,0.5}
\definecolor{mauve}{rgb}{0.58,0,0.82}

\usetikzlibrary {
  matrix,
  shapes,
  arrows,
  shadows,
  calc,
  chains,
  decorations.pathmorphing,
  decorations.text,
  arrows.meta,
  patterns,
  fit,
  bending,
  shadows,
  shapes.geometric,
  snakes,
  intersections
}

\tikzset{
  label/.style={
    font=\scriptsize\itshape,
    inner sep=0.4em
  },
  point/.style={
    circle,
    fill=black,
    text width=0.3em,
    inner sep=0
  },
  state/.style={
    circle,
    text width=0.8em,
    inner sep=0.1em,
    text depth=0.08em,
    draw,
    font=\scriptsize
  },
  participant/.style={
    draw=black,
    rounded corners,
    semithick,
    font=\footnotesize,
    text height=0.2cm,
text centered,
    anchor=base
  },
  specification/.style={
    font=\small,
    semithick,
    draw=black
  },
  synthesiser/.style={
    draw=black,
    dotted,
    semithick,
    inner sep=2pt,
    font=\footnotesize
  },
  monitor/.style={
draw=black,
    fill=white,
    semithick,
    minimum width=0.5cm,
    minimum height=0.5cm,
    font=\footnotesize,
    text height=0.2cm,
    drop shadow=ashadow
  },
  point/.style={
    minimum size=0pt, 
    inner sep=0pt
  },
  ltspoint/.style={
    circle,
    fill=black,
    draw=white,
    line width=0.5mm,
    text width=0.25em,
    inner sep=0,
    font=\footnotesize
  },
  ashadow/.style={
    opacity=.3, 
    shadow xshift=0.6mm, 
    shadow yshift=-0.6mm 
  },
  oppashadow/.style={
    opacity=.3, 
    shadow xshift=-0.6mm, 
    shadow yshift=-0.6mm 
  },
  processDiagram/.style={
    draw=black,
    fill=white,
    semithick,
    minimum width=0.5cm,
    minimum height=0.5cm,
    font=\footnotesize,
    text height=0.2cm,
    drop shadow=ashadow
  },
  event/.style={
    draw=black,
    semithick,
    minimum width=0.4cm,
    minimum height=0.4cm,
    font=\scriptsize
  },
  trap/.style={
    trapezium, 
    trapezium angle=67.5, 
    draw,
    inner ysep=5pt, 
    outer sep=0pt,
    minimum height=1.81mm, 
    minimum width=0pt
  },
  static-process/.style={
    draw=black,
    fill=white,
    semithick,
    minimum width=0.5cm,
    minimum height=0.5cm,
    font=\footnotesize,
    text height=0.2cm
  },
  pill/.style={
    draw,
text=black,
    fill=white,
    minimum width=0.84em, 
    minimum height=0.8em,
    inner sep=0.1em,
    outer sep=0,
    font=\scriptsize\sffamily,
    align=center,
    rounded corners=0.4em  
  }
}

\pgfdeclaredecoration{penciline}{initial}{
    \state{initial}[width=+\pgfdecoratedinputsegmentremainingdistance,auto corner on length=1mm,]{
        \pgfpathcurveto {\pgfqpoint{\pgfdecoratedinputsegmentremainingdistance}
                            {\pgfdecorationsegmentamplitude}
        }
        {\pgfmathrand
        \pgfpointadd{\pgfqpoint{\pgfdecoratedinputsegmentremainingdistance}{0pt}}
                        {\pgfqpoint{-\pgfdecorationsegmentaspect\pgfdecoratedinputsegmentremainingdistance}{\pgfmathresult\pgfdecorationsegmentamplitude}
                        }
        }
        {\pgfpointadd{\pgfpointdecoratedinputsegmentlast}{\pgfpoint{1pt}{1pt}}
        }
    }
    \state{final}{}
} 
\usepackage{xargs}
\usepackage[normalem]{ulem} 

\usepackage{xifthen}        \newcommand{\ifempty}[3]{\ifthenelse{\isempty{#1}}{#2}{#3}}

\DeclareGraphicsExtensions{.png,.PNG,.pdf,.PDF,.jpg,.mps,.jpeg,.jbig2,.jb2,.JPG,.JPEG,.JBIG2,.JB2
}

\newif\ifemi

\newcommandx{\preprint}[3][1=preprint,2=Springer]{
  \ifempty{#1}{}{
	 \ \\[1em]\noindent
	 \textbf{Disclaimer}
    The published version of this paper is~\cite{#1} (\copyright\ #2).
  }
}

\usepackage[disable]{todonotes}

\usepackage[most]{tcolorbox}
\newtcolorbox{markbox}{
  enhanced,
  breakable,
  size=minimal,
  parbox=false,
  after={\par},
  before upper={\indent},
  colback=white,
  overlay = {
	 \draw[line width=2pt]
	 (frame.north east) -| ([xshift=3mm]frame.east) |-(frame.south east);
  },
  overlay first={\draw[line width=2pt] (frame.north east) -| ([xshift=3mm]frame.south east);},
  overlay middle={\draw[line width=2pt] ([xshift=3mm]frame.north east) -- ([xshift=3mm]frame.south east);},
  overlay last={\draw[line width=2pt] ([xshift=3mm]frame.north east) |- (frame.south east);},
}

\newcommand{\eMcomm}[2][check]{\ifthenelse{\equal{#1}{new}}{{\color{red}#2}}{\ifthenelse{\equal{#1}{changed}}{{\color{teal}{#2}}}{\todo[color=orange!20]{\tiny eM: \color{NavyBlue}#1}{\color{black}{#2}}}}}

\newcommand{\hidden}[1]{}
\newcommand{\hide}[1]{}

\newcommand{\cf}[2]{
  \fontsize{#1}{#1}{\selectfont{#2}}
}

\makeatletter
\newcommand{\dolist}[2]{\def\nextitem{\def\nextitem{#1}}\@for \el:=#2\do{\nextitem\textbf{\el}}}

\def\mktest#1{
  \def\transform##1+##2+##3{##1 piu' ##2 * ##3\penalty0}\do{\expandafter\transform#1}}
\def\mksubscript#1{
  \def\transform##1[##2]{##1_{##2}\penalty0}\do{\expandafter\transform#1}}
\newcommand{\mapcmd}[3][{, }]{\def\nextitem{\def\nextitem{#1}}\@for \el:=#3\do{\nextitem{#2{\el}}}}
\makeatother

\ifemi
\usepackage{showlabels}

\newcommand{\emi}[2]{
  \marginpar{\fcolorbox{red}{shadecolor}{\cf{#1}{{#2}}}}
}
\newcommand{\emic}[2]{\par
  \fcolorbox{red}{shadecolor}{\parbox{\linewidth}{ 
      \color{gray}
      \begin{description}
      \item[{\color{blue} #2}]{\sf #1}
      \end{description}}}
}
\else
\newcommand{\emi}[2]{}
\newcommand{\emic}[2]{}{}
\fi

\newcommand{\noarg}{}
\newcommand{\mkfun}[4][\colorFun]{
  \newcommand{#2}[1][#4]{
    {#1\ensuremath{\mathsf{#3}}}
    \ifempty{##1}{\noarg}{
      ({##1})}
  }
}

\mkfun{\head}{hd}{}
\mkfun{\tail}{tl}{}

\newcommand{\sst}{\;\big|\;}
\newcommand{\qst}{\;\colon\;}

\newcommand{\bnfdef}{\ ::=\ }

\newcommand{\bnfmid}{\;\ \big|\ \;}

\newcommand{\ie}{\text{i.e.,}\xspace}

\newcommand{\eg}{\text{e.g.,}\xspace}

{\bfseries}{\rmfamily}

\makeatletter
\newcommand*{\da@rightarrow}{\mathchar"0\hexnumber@\symAMSa 4B }
\newcommand*{\da@leftarrow}{\mathchar"0\hexnumber@\symAMSa 4C }
\newcommand*{\xdashrightarrow}[2][]{\mathrel{\mathpalette{\da@xarrow{#1}{#2}{}\da@rightarrow{\,}{}}{}}}
\newcommand{\xdashleftarrow}[2][]{\mathrel{\mathpalette{\da@xarrow{#1}{#2}\da@leftarrow{}{}{\,}}{}}}
\newcommand*{\da@xarrow}[7]{\sbox0{$\ifx#7\scriptstyle\scriptscriptstyle\else\scriptstyle\fi#5#1#6\m@th$}\sbox2{$\ifx#7\scriptstyle\scriptscriptstyle\else\scriptstyle\fi#5#2#6\m@th$}\sbox4{$#7\dabar@\m@th$}\dimen@=\wd0 \ifdim\wd2 >\dimen@
    \dimen@=\wd2 \fi
  \count@=2 \def\da@bars{\dabar@\dabar@}\@whiledim\count@\wd4<\dimen@\do{\advance\count@\@ne
    \expandafter\def\expandafter\da@bars\expandafter{\da@bars
      \dabar@ 
    }}\mathrel{#3}\mathrel{\mathop{\da@bars}\limits
    \ifx\\#1\\\else
      _{\copy0}\fi
    \ifx\\#2\\\else
      ^{\copy2}\fi
  }\mathrel{#4}}
\makeatother

\newcommand{\quo}[1]{\lq\lq {#1}\rq\rq}
\def\finex{{\unskip\nobreak\hfil
\penalty50\hskip1em\null\nobreak\hfil$\diamond$
\parfillskip=0pt\finalhyphendemerits=0\endgraf}}

\newcommand{\asort}[1][s]{\ensuremath{\mathtt{#1}}}

\definecolor{shadecolor}{rgb}{1,0.99,0.9}
\definecolor{bg}{rgb}{0.95,0.95,0.95}

\newcommand{\abcattr}[1][a]{\textsf{#1}}
\newcommand{\abccond}[1][\rho]{#1}

\def\colorExp{\color{NavyBlue}}

\newcommand{\abcexp}[1][e]{\colorExp #1}
\newcommandx{\abctuple}[1][1 = t]{\llparenthesis{#1}\rrparenthesis}
\newcommandx{\abcget}[2][1=a,2={id},usedefault=@]{
  \ptp[{#1}]{\colorOp .}\abcattr[{#2}]
}
\newcommandx{\abcptp}[2][1=a,2=\abccond,usedefault=@]{\ifempty{#1}{}{\ptp[{#1}] \ifempty{#2}{}{{\colorOp \shortmid}}} {#2}}
\newcommandx{\abcint}[6][1=a,2=\abccond,3=e,4=e',5=b,6=\abccond',usedefault=@]{
  \abcptp[{#1}][{#2}]
  \ {\colorOp \xrightarrow{\scriptstyle \abcexp[#3]\quad\abcexp[#4]}}\ 
  \abcptp[{#5}][{#6}]
}
\newcommandx{\mkabcint}[8][3=a,4=\abccond,5=e,6=e',7=b,8=\abccond',usedefault=@]{
  \node[bblock, #1] (#2) {$\abcint[{#3}][{#4}][{#5}][{#6}][{#7}][{#8}]$};
}

\tikzset{
    abccallout/.style={
      fill=green!10,
		opacity=.5,
		overlay,
		align=center,
      cloud callout,
		cloud puffs=15,
		aspect=2.5,
		cloud ignores aspect,
		cloud puff arc=100,
		shading=ball
    }
  }

\newcommandx{\abcP}[6][1=P,2=K,3=.1cm,4=1cm,5=north east,6=proc,usedefault=@]{
  \begin{tikzpicture}
	 \node[fill=blue!10, shape=circle] (#6) {$\p[#1]$};
	 \node[abccallout, above = #3 of #6, xshift=#4, callout absolute pointer={(#6.#5)}] {$#2$}
	 ;	 
	 \draw[decorate,decoration={expanding waves,angle=7,segment length = .05cm}] (#6.east) -- ++(.5cm,0)
	 ;
  \end{tikzpicture}
}

\ExplSyntaxOn
\NewDocumentCommand{\ucgreek}{m}
 {
  \str_case:nn { #1 }
   {
    {A}{\mathrm{A}}
    {B}{\mathrm{B}}
    {C}{\Sigma}
    {D}{\Delta}
    {E}{\mathrm{E}}
    {F}{\Phi}
    {G}{\Gamma}
    {H}{\mathrm{H}}
    {I}{\mathrm{I}}
    {J}{\Theta}
    {K}{\mathrm{K}}
    {L}{\Lambda}
    {M}{\mathrm{M}}
    {N}{\mathrm{N}}
    {O}{\mathrm{O}}
    {P}{\Pi}
    {Q}{\mathrm{X}}
    {R}{\mathrm{P}}
    {S}{\Sigma}
    {T}{\mathrm{T}}
    {U}{\Upsilon}
{W}{\Omega}
    {X}{\Xi}
    {Y}{\Psi}
    {Z}{\mathrm{Z}}
   }
 }
\NewDocumentCommand{\lcgreek}{m}
 {
  \str_case:nn { #1 }
   {
    {a}{\alpha}
    {b}{\beta}
    {c}{\varsigma}
    {d}{\delta}
    {e}{\varepsilon}
    {f}{\varphi}
    {g}{\gamma}
    {h}{\eta}
    {i}{\iota}
    {j}{\vartheta}
    {k}{\kappa}
    {l}{\lambda}
    {m}{\mu}
    {n}{\nu}
    {o}{o}
    {p}{\pi}
    {q}{\chi}
    {r}{\rho}
    {s}{\sigma}
    {t}{\tau}
    {u}{\upsilon}
{w}{\omega}
    {x}{\xi}
    {y}{\psi}
    {z}{\zeta}
   }
 }
\ExplSyntaxOff

\def\pctcolor{OliveGreen}

\newcommand{\dslsep}[1]{{\color{\pctcolor}\ensuremath{#1}}}

\makeatletter
\newcommand{\dslbrc}[1]{\ensuremath{
	 \ifthenelse{\equal{#1}{...}}{\ensuremath{\cdots}}{
		\def\@p##1->##2{{\it ##1} \dslsep\to {\it ##2}}
		\do{\expandafter\@p#1}
	 }
  }
}
\makeatother

\usepackage{mdframed}

\makeatletter
\begin{document}

\title{\tool: Connected OpenAPI Test Synthesis for RESTful Applications}

\author{Christian Bartolo Burl\`{o}\inst{1}\orcidID{0000-0002-0016-086X} \and
  \\
  Adrian Francalanza\inst{2}\orcidID{0000-0003-3829-7391} \and
  \\
  Alceste Scalas\inst{3}\orcidID{0000-0002-1153-6164} \and
  \\
  Emilio Tuosto\inst{1}\orcidID{0000-0002-7032-3281}
}

\authorrunning{C. Bartolo Burl\`{o} \etal}

\institute{Gran Sasso Science Institute, L'Aquila \and Department of Computer Science, University of Malta, Msida, Malta \and DTU Compute, Technical University of Denmark, Kongens Lyngby, Denmark}

\Crefname{section}{\S}{Sections}
\Crefname{figure}{Fig.\@}{Figures}
\Crefname{definition}{Def.\@}{Definitions}

\maketitle

\begin{abstract}
  We present a novel model-driven approach for testing RESTful
  applications.
We introduce a ($i$) domain-specific language for OpenAPI
  specifications and ($ii$) a tool to support our methodology.
Our DSL, called \model, is inspired by session types and enables the
  modelling of communication protocols between a REST client and
  server.
Our tool, dubbed \tool, generates (randomised) model-based test
  executions and reports software defects.
We evaluate the effectiveness of our approach by applying it to test
  several open source applications.
Our findings indicate that our methodology can identify nuanced
  defects in REST APIs and achieve comparable or superior code
  coverage when compared to much larger handcrafted test suites.

\medskip
\emph{\textbf{DISCLAIMER.} This preprint is the author version of the paper published at COORDINATION 2024:}\\%
\url{https://doi.org/10.1007/978-3-031-62697-5_5}
\end{abstract}
 
\section{Introduction}\label{sec:introduction}

Modern software is increasingly composed of concurrent and distributed components that are independently developed, and function by exchanging data across a communication network.
The interaction of such components may be based on classic client-server Internet protocols (such as SMTP, IMAP, and POP3), various forms of remote procedure calls (RPCs),  web-based standards --- such as the REpresentational State Transfer (REST) architectural design.

Ensuring the correctness and reliability of these applications is notoriously hard. Software developers typically create handcrafted test suites, which must address many (and ever-growing) application usage scenarios: 
It is often the case that such manually-written test suites are developed intermittently, over long periods of time, by a variety of testers. 
This makes the software development and testing process error-prone and susceptible to inconsistencies~\cite{DBLP:conf/kbse/StallenbergOP21,DBLP:journals/software/LenarduzziDMPT21}. 
A number of efforts have emerged \eMcomm[aimed at]{with the aim of} simplifying the testing process of component based-software by streamlining the amount of manual work through automatic test generation. 
Most of these efforts focus on automatic test generation for applications exposing an API over a network, where there is a considerable interest in addressing the popular REST API style~\cite{DBLP:conf/edoc/Ed-DouibiIC18, DBLP:conf/kbse/StallenbergOP21, DBLP:journals/tse/SeguraPTC18, DBLP:journals/ese/ZhangMA21, DBLP:conf/icst/ViglianisiDC20, DBLP:conf/icse/AtlidakisGP19, DBLP:conf/qrs/Arcuri17, DBLP:conf/icst/KarlssonCS20, Pinheiro2013ModelBasedTO, testtherest, DBLP:conf/erlang/SeijasLT13, DBLP:conf/www/FertigB15}.

\begin{example}\label{eg:intro-example}
  \eMcomm[{\Cref{lst:openapiSpec} shows a fragment of a REST API offered by a web application:
  the \emph{SockShop} microservice demo application \cite{sockshop}.
  The API fragment is written as an OpenAPI specification \cite{openAPI}.}]{
\Cref{lst:openapiSpec} shows a fragment of a REST API of the
  \emph{SockShop} web application~\cite{sockshop} written as
  an OpenAPI specification~\cite{openAPI}.
}
A new customer can be created by passing the necessary credentials as
payload to the operation with ID \texttt{addCust} (\eMcomm[line
3]{lines 3-4}); this operation returns a unique reference identifier
for the customer, meant to be used in subsequent operations\eMcomm[.
The API also offers operations]{} to create a payment card (\texttt{addCard}) and an address (\texttt{addAddr}) to associate with (previously created) customers.
The customer information  including any cards and addresses linked to it can be retrieved with \texttt{getCust} and deleted with \texttt{deleteCust}.
\qed
\end{example}

\begin{figure}[!t]
\begin{subfigure}{0.45\textwidth}
  \small\begin{Verbatim}[commandchars=\\\{\},numbers=left,numbersep=1mm,frame=single,breaklines]
  paths:
    /customer: 
      POST:
        operationId: addCust
        responses: '201' ...
    /customer/{id}:
      GET:
        operationId: getCust
        responses: '200' ...
      DELETE:
        operationId: deleteCust
        responses: '204' ...
    /card:
      POST:
        operationId: addCard
        responses: '201' ...
    /address:
      POST:
        operationId: addAddr
        responses: '201' ...
  \end{Verbatim}
  \caption{OpenAPI spec. for an online store.\label{lst:openapiSpec}}
\end{subfigure}
\hfill
\begin{subfigure}{0.45\textwidth}
  \scalebox{1}{\begin{tikzpicture}
  \node (client) [participant]{\small{client}};
  \node (server) [participant, right=3cm of client]{\small{server}};

  \node (client-end) [point, below=4.5cm of client]{};
  \node (server-end) [point, below=4.5cm of server]{};

  \draw[-open square, semithick] (client) edge (client-end);
  \draw[-open square, semithick] (server) edge (server-end);

  \node (client-addCust) [point, below=0.5cm of client]{};
  \node (server-addCust) [point, below=0.5cm of server]{};
  
  \draw[semithick,->] (client-addCust) edge node[above,yshift=-2pt]{\small\lab{addCust}(\asort[Customer])}  (server-addCust);

  \node (client-C200_1) [point, below=0.5cm of client-addCust]{};
  \node (server-C200_1) [point, below=0.5cm of server-addCust]{};

  \draw[semithick,->] (server-C200_1) edge node[above,yshift=-2pt]{\small\lab{C201}(\asort[custID])} (client-C200_1);

  \node (client-addCard) [point, below=0.5cm of client-C200_1]{};
  \node (client-addCard-left) [fill=black, draw=black, minimum size=0.2pt, inner sep=0pt, left=0.3cm of client-addCard]{};
  \draw[dashed,-] (client-addCard-left) edge (client-addCard.east);
  \node (server-addCard) [point, below=0.5cm of server-C200_1]{};
  \draw[semithick,->] (client-addCard) edge node[above,yshift=-2pt]{\small\lab{getCust}(\asort[custID])} (server-addCard);

  \node (client-C200_2) [point, below=0.5cm of client-addCard]{};
  \node (server-C200_2) [point, below=0.5cm of server-addCard]{};
  \node (server-C200_2-right) [fill=black, draw=black, minimum size=0.2pt, inner sep=0pt, right=0.3cm of server-C200_2]{};
  \draw[dashed,-] (server-C200_2) edge (server-C200_2-right);

  \draw[semithick,->] (server-C200_2) edge node[above,yshift=-2pt]{\small\lab{C200}(\asort[Customer])} node[below,align=center, yshift=0.2cm]{$\mathsmaller{\vdots}$} (client-C200_2);

  \node (client-delCust) [point, below=1cm of client-C200_2]{};
  \node (server-delCust) [point, below=1cm of server-C200_2]{};
  \node (client-delCust-left) [fill=black, draw=black, minimum size=0.2pt, inner sep=0pt, left=0.3cm of client-delCust]{};
  \draw[dashed,-] (client-delCust-left) edge (client-delCust);
  \draw[semithick,->] (client-delCust) edge node[above,yshift=-2pt]{\small\lab{delCust}(\asort[custID])} (server-delCust);

  \node (client-C200_3) [point, below=0.5cm of client-delCust]{};
  \node (server-C200_3) [point, below=0.5cm of server-delCust]{};

  \draw[semithick,->] (server-C200_3) edge node[above,yshift=-2pt]{\small\lab{204}}  (client-C200_3);

  \node (client-400) [point, below=0.5cm of client-C200_3]{};
  \node (server-400) [point, below=0.5cm of server-C200_3]{};
  \draw[semithick,->] (server-400) edge node[above,yshift=-2pt]{\small\lab{400}}  (client-400);

  \node (server-recurse-point) [point, below=0.25cm of server-C200_1]{};
  \node (server-recurse-point-right) [fill=black, draw=black, minimum size=0.2pt, inner sep=0pt, right=0.3cm of server-recurse-point]{};
  \draw[dashed,<-] (server-recurse-point) edge (server-recurse-point-right);

  \node (server-options-point) [point, below=0.25cm of server-C200_2]{};
  \node (server-options-point-right) [fill=black, draw=black, minimum size=0.2pt, inner sep=0pt, right=0.3cm of server-options-point]{};
  \draw[dashed,-] (server-options-point) edge (server-options-point-right);

  \draw[dashed,-] (server-options-point-right) edge node[right,above,rotate=-90,yshift=-2pt] {\small\textit{recursion}} (server-recurse-point-right);

  \node (server-201-point-right) [fill=black, draw=black, minimum size=0.2pt, inner sep=0pt, right=0.7cm of server-C200_1]{};
  \node (server-400-point-right) [fill=black, draw=black, minimum size=0.2pt, inner sep=0pt, right=0.7cm of server-400]{};
  \draw[dashed,-] (server-400) edge (server-400-point-right);
  \draw[dashed,-] (server-C200_1) edge (server-201-point-right);
  \draw[dashed,-] (server-201-point-right) edge node[right,above,rotate=-90,yshift=-2pt] {\small\textit{choice}} (server-400-point-right);

  \node (client-choice-point) [point, below=0.25cm of client-C200_2]{};
  \node (client-choice-point-right) [fill=black, draw=black, minimum size=0.2pt, inner sep=0pt, left=0.3cm of client-choice-point]{};
  \draw[dashed,-] (client-choice-point) edge (client-choice-point-right);

  \draw[dashed,-] (client-addCard-left) edge node[left,above,rotate=90,yshift=-2pt] {\small\textit{choice}} (client-delCust-left);

  \node (client-1) [fill=white, minimum size=1pt, inner sep=3pt, below=0.4cm of client-C200_2]{.};
  \node (server-1) [fill=white, minimum size=1pt, inner sep=3pt, below=0.4cm of server-C200_2]{.};
\end{tikzpicture} }
  \caption{Message Sequence Chart outlining the invocation dependencies  in \Cref{lst:openapiSpec}.}\label{fig:msc}
\end{subfigure}
\caption{Typical documentation provided for REST APIs.\label{fig:documentation}}
\end{figure}
The OpenAPI \eMcomm[specification snippet from]{in} \Cref{lst:openapiSpec} is a typical instance of a REST API description. 
OpenAPI is 
used pervasively for such specifications: it provides 
information on the available URIs, the corresponding HTTP methods and respective responses with the associated data formats (omitted from \Cref{lst:openapiSpec}). 
However, OpenAPI does not 
specify the
relationships and dependencies between different API invocations.
For instance, \Cref{lst:openapiSpec} does \emph{not} express that:
\begin{itemize}
\item Customer information  can be added (\texttt{addCard}, \texttt{addAddr}), retrieved (\texttt{getCust}), or deleted (\texttt{deleteCust}) only after the customer is created (\texttt{addCust}).
\item Operations \texttt{addCard}, \texttt{addAddr} or \texttt{getCust} cannot be executed after the customer has been deleted (\texttt{deleteCust}).
\item Operations \texttt{addCard}, \texttt{addAddr} and \texttt{getCust} may be interleaved without effecting the successful outcome of the other invocations.
\item The variable \texttt{custId}, an ID given to the customer upon creation is to be used to perform any operation on the customer (such as retrieval or deletion).
\end{itemize}

These dependencies induce an ordering in the API invocations, depicted in the message sequence chart in \Cref{fig:msc}.
Unfortunately, this information is often omitted (cf.~\cite{petclinic,users-regsitry,sockshop}) or only informally stated (as \eg in~\cite{features-service}). 
In development contexts, this is often sufficient as developers are typically able to deduce the intended order of invocation by examining the descriptions of requests and the data types exchanged by each operation.
However, this omission becomes problematic 
when testing.
Deviations from the dependencies identified earlier in \Cref{fig:msc} can be categorized as \emph{logic-based faults}, that is, errors stemming from incorrect or unintended actions within the business logic of the application.
Such violations are often intricate and context-specific, arising from the unique interactions and workflows within the application. 
These faults are not easily detectable through basic system responses or standard error codes and typically necessitate comprehensive, scenario-based testing for identification.
Furthermore, these faults are intrinsically linked to the essential operations and 
branching mechanisms of the application, impacting its ability to perform as intended.

Previous work on REST API testing proposes various \emph{fully-automated} test generation strategies \cite{DBLP:conf/gecco/ZhangMA19,
  DBLP:conf/icse/AtlidakisGP19, DBLP:conf/icst/KarlssonCS20,
  DBLP:conf/kbse/StallenbergOP21}. 
These \emph{push-button} tools
excel at detecting \emph{systemic errors} that are related to the system's general functioning such as server crashes, malformed requests, unauthorized access attempts, or resource not found errors.
However, it is unlikely that logic-based errors are uncovered by invoking the API without an explicit pre-defined model of the intended interaction with the API. 
\Eg in the \emph{SockShop} application from \Cref{eg:intro-example}, such tools are able to detect if the \sut crashes when 
creating a new customer, but it is unlikely that they are able to detect whether the \sut allows the retrieval of a previously deleted customer.   
\eMcomm[To detect such errors, there typically has to be a manual test
that creates a customer, deletes it and then try to retrieve it (using
the previous data).]{A test to detect such errors should create a
  customer and then try to retrieve the same customer after having
  deleted it.
}
It has been empirically shown that the quality of tests substantially improves when conducted in a \emph{stateful} manner, \ie by exploring states of the system under test (\sut) that are 
reachable via sequences of 
invocations \cite{DBLP:conf/icse/AtlidakisGP19}.
A recent work concludes that: \emph{\quo{existing tools fail to achieve high code coverage due to limitations of the approaches they use for generating parameter values and detecting operation dependencies}} \cite{kim2022automated}.

The state-of-the-art of testing REST APIs can be 
seen on a spectrum.
At one end, manual tests, though labor-intensive, excel at uncovering complex, logic-based errors.
At the opposite end are fully-automated testing tools, which, while requiring minimal effort, fall short in detecting 
eloaborate errors.
This work seeks a middle ground, leveraging a \emph{model-based} approach. 
Our goal is to automate the identification of more 
logic-based faults, 
thus bridging the gap between labor-intensive manual testing and the scope of fully-automated tools.

\paragraph*{Contributions.} We present a model-based, tool-supported methodology for the automatic testing of web applications exposing REST APIs.
Our 
approach
allows the specification of API dependencies and constraints. 
Our contributions are:
\begin{enumerate}[nosep]
\item \model: a domain-specific language designed to specify dependencies between requests in an OpenAPI specification
that capture the state of the interaction with the \sut and the sequencing of message exchanges;
\item \tool: an automated tool 
leveraging \model models to generate tests that interact with the \sut and  
assess the correctness of its responses.
\item Experimental results 
demonstrating the effectiveness of \model and \tool
using 
case studies from various application domains. We compare our results 
against RESTful API testing tools and manually-written tests.
\end{enumerate}
The key benefits of our approach are in terms of ($i$) a high degree of expressiveness due to the possibility of specifying data dependencies, ($ii$) effectiveness of the approach that can identify logic-based faults, and ($iii$) high level of coverage.
\emph{\textbf{We will submit \tool for evaluation as companion artefact of this paper, allowing for the reproduction of our results.}}

\section{A Model-Driven, DSL-Based Methodology}
\label{sec:aug-session-types}

\subsection{Background}

Based on the REST architectural style \cite{rest-thesis} and HTTP protocol, REST APIs facilitate software services on the web by exposing \emph{resources} through HTTP URIs, which are manipulated using standard HTTP \emph{request methods} like \texttt{GET}, \texttt{POST}, \texttt{PUT}, and \texttt{DELETE}. Responses are represented by standard HTTP codes (\eg 200 for success, 404 for missing resources, 500 for server errors). While OpenAPI and GraphQL~\cite{graphql} are prominent for documenting REST APIs, our emphasis on OpenAPI stems from its broad adoption and tool support for generating client libraries across various programming languages. Though our methodology primarily targets OpenAPI, it's flexible for other web API standards like GraphQL.

The technical nature of REST APIs, their complex state-dependent interactions, and their evolving nature introduces challenges for which existing model-based testing methods and tools are ill-suited (see related work in \Cref{sec:related-work}).
To bridge this gap, we introduce the \model DSL (\Cref{sec:copenapi}), designed to articulate the intricate sequences and state dependencies characteristic of RESTful interactions, enabling precise testing beyond the capabilities of existing models and tools.

\subsection{Methodology}

\begin{figure}[tp]
  \centering
  \scalebox{0.8}{\scalebox{1.3}{\begin{tikzpicture}[decoration=penciline]
\node (sessiontype) [decorate, draw] {\footnotesize\textit{Model}};
  \node (driver) [draw=black, processDiagram, align=center, below=0.4cm of sessiontype] {\footnotesize\textit{Test driver}};
  \node (logs) [participant, draw=black, align=center, minimum height=0.45cm, below=0.4cm of driver, xshift = -0.5cm] {\scriptsize\texttt{Logs}};
  \node (tests) [participant, draw=black, align=center, minimum height=0.45cm, below=0.4cm of driver, xshift = 0.5cm] {\scriptsize\texttt{Tests}};
  \node (sut) [processDiagram, draw=black, fill=gray!30, text=black, align=center, right=0.5cm of driver] {\footnotesize\textit{\sut}};

  \node (testing)[left=0.7cm of driver] {\footnotesize\textit{Testing}};
  \node (design)[above=0.4cm of testing] {\footnotesize\textit{Modelling}};
  \node (analysis)[below=0.4cm of testing] {\footnotesize\textit{Analysis}};
  
  \draw[dashed] (-0.9,-0.4) -- (1.8,-0.4);
  \draw[dashed] (-0.9,-1.3) -- (1.8,-1.3);

\draw[<->] (driver) edge (sut);
  \draw[-stealth] (sessiontype) edge (driver);
\draw[-stealth] (driver) -- (logs);
  \draw[-stealth] (driver) -- (tests);
  \draw[-stealth] (analysis.west) -| (design.west); 
\end{tikzpicture}
}%
 }
  \caption{Methodology overview}
  \label{fig:methodology}
 \vspace{-0.5cm}
\end{figure}

Our methodology consists of the three phases
depicted in \Cref{fig:methodology}:

\begin{enumerate}[label={\textbf{\textsf{(P\arabic*)}}}]
\item\label{item:method:model} \emph{modelling phase:} construct a model that describes how a client application might interact with the \sut, and how the \sut is expected to respond to the client's inputs;
\item\label{item:method:test} \emph{testing phase:} automatically generate a \emph{test driver} that tests the \sut by interacting with it according to the model, and reports whether the \sut responses violate the model;
\item\label{item:method:analysis} \emph{analysis phase:} inspect the outputs produced by test driver to identify faults in the \sut.
\end{enumerate}
The phases \ref{item:method:model}, \ref{item:method:test} and
\ref{item:method:analysis} are \eMcomm[are iterative: the model can be
refined and enriched based on the findings of the analysis;
furthermore, the model can evolve during the life cycle of the
\sut.]{iterated when the model has to be refined according to the
  findings of the analysis or it has to be evolved according to the
  life cycle of the \sut.}
The main goal of our approach is to automate the
testing phase, and let testers shift their efforts to the modelling
phase, reducing the need to develop (and maintain) a large suite of
handcrafted test.

The first step towards the methodology entails determining a suitable
model for phase \ref{item:method:model}, allowing for the automatic
derivation of tests for RESTful applications. To write such model,
\eMcomm[we develop a domain-specific language (DSL) based on
OpenAPI. Called \model, it has its syntax inspired by \emph{session
  types} and is presented in \Cref{sec:copenapi}.]{we design \model
  (see \Cref{sec:copenapi}) taking inspiration from \emph{(binary)
	 session types}~\cite{DBLP:journals/csur/HuttelLVCCDMPRT16}.}

The next step in our methodology is \ref{item:method:test}, \ie the testing phase. In this phase, the test driver interacts with the \sut by sending and receiving messages according to the \model model. 
The test driver needs to perform two types of interactions with the \sut:
\begin{itemize}
  \item the test driver must send requests to the \sut of the correct format and payload type; and
\item the test driver needs to check that the responses received from the \sut do not violate the \model model. These violations may be of three forms: 
  \begin{itemize}[nosep]
    \item the response code received is not valid, or
    \item the payload data does not have the expected format, or
    \item the response code and payload data format are valid, but the payload
      data causes an assertion violation in the model (\eg by violating a constraint
      involving data from a previous request/response).
  \end{itemize}
\end{itemize}

The last phase of the methodology \Cref{item:method:analysis} is devoted to the identification of
faults in the \sut, by inspecting the outputs of the test driver. 
The test driver produces a successful test when it manages to complete a full traversal of the \model model without finding any of the aforementioned errors. 
To help with the analysis, the test driver produces two outputs:
\begin{enumerate}
\item a log file with information about every performed test: the
  random seed used to generate it, the sequence of messages sent and
  received and their respective payloads, and the test outcome
  (pass/fail); and
\item an offline representation of failed tests, usable for manually
  reproducing faults of the \sut without re-executing the test driver, useful for, \eg bug reporting.
\end{enumerate}

\section{A DSL for OpenAPI}\label{sec:copenapi}

We introduce a DSL based on an augmented form of \emph{session types} \cite{DBLP:journals/csur/HuttelLVCCDMPRT16,dd09,vas12}. 
Intuitively, a session type specifies the valid sequences of message exchanges (\ie the \emph{protocol}) that regulate the interaction between communication programs; here we focus on \emph{binary session types}, which model how a program is expected to interact with just one other program (in our case, a client with a RESTful server). 

The formal syntax of our augmented session types $\stS$ is:
\eMcomm[the subscripts $i$ are messy...introduce a notation such as $\overline{x:\stT}$ for sequences of declarations?]{}
\[
\begin{array}{rcll}
  \stS & \bnfdef & \stSeli{i}{I}[\stTuple][\textsf{gen}_i] & \textit{(internal choice)} \\[0.5mm]
       & \bnfmid & \stBrai{i}{I}[\stTuple][\textsf{dec}_i] & \textit{(external choice)} \\[0.5mm]
       & \bnfmid & \stRec{X}.\stS & \textit{(recursion)} \\[0.5mm]
       & \bnfmid & \stRecVar{X} & \textit{(recursion variable)} \\[0.5mm]
       & \bnfmid & \stEnd & \textit{(termination)}
  \\[2mm]
  \stT & \bnfdef & \texttt{Int} \bnfmid \texttt{String} \bnfmid \ldots & \textit{(data types)}
\\[2mm]
  \stG & \bnfdef & \texttt{Int}(g) \bnfmid \texttt{String}(g) \bnfmid \ldots & \textit{(data types with generators)}
\end{array}
\]
where $I$ is a finite non-empty set of indexes and for $i \in I$,
$\textsf{gen}_i$ is a \emph{generator} assignment
$x_{i,1}:\textsf{G}_{i,1},\ldots,x_{i,n_i}:\textsf{G}_{i,n_i}$
(cf. \cref{ex:gen}) and $\textsf{dec}$ is a type assignment
$x_{i,1}:\textsf{T}_{i,1},\ldots,x_{i,n_i}:\textsf{T}_{i,n_i}$) with
$x_{i,1}, \ldots, x_{i,n_i}$ pairwise distinct.\footnote{We use
  standard data types, including standard types (\eg \texttt{Int},
  \texttt{String}, ...) and user-defined types.
Generators are also user-defined, tailored to the particular request
  being sent.
}

The communication units of our syntax are the \emph{output} and \emph{input} prefixes, respectively \stOutOp\textit{operationId} and \stInOp\textit{responseCode} where \textit{operationId} is the corresponding identifier referring to the specific request in the OpenAPI specification and \textit{responseCode} is the HTTP response code that is returned from the service after the request is made (see \Cref{sec:related-work}).
Assertions $A$ are boolean expressions that may refer to the variables occurring in preceding payload descriptions.

We illustrate the usage of \model with the following example which elaborates on the protocol in \Cref{fig:msc}.

\begin{figure}
  \small\begin{Verbatim}[commandchars=\\\{\},numbers=left,numbersep=1mm,frame=single,breaklines]
S_shop = !addCust(apiKey: String(genApiKey), c1: Customer(genCustInfo)).
         ?C201(custId: String).
         rec X.(
           +\{ !addCard(apiKey, custId, card: Card(getCardInfo)) . 
                   ?C201(CardId: String) . X,
               !addAddr(apiKey, custId, addr: Address(getAddrInfo)) . 
                   ?C201(addressId: String) . X,
               !getCust(apiKey, custId) . 
                   ?C200(c2: Customer)<checkCustomer(c1,c2)> . X,
               !deleteCust(apiKey, custId).?C204() . end
            \}
         )
    \end{Verbatim}
    \vspace{-3mm}
  \caption{Example of a \model model using the OpenAPI specification in \Cref{lst:openapiSpec}, formalising the message-sequence diagram in \Cref{fig:msc}.}
\label{fig:model-example}
\end{figure}

\begin{example}\label{ex:gen}
  \Cref{fig:model-example} shows a testing model written in \model. 
The model corresponds to the message-sequence diagram from \Cref{fig:msc}.
The model specifies that the interaction protocol should first create a customer by invoking \lab{addCust} together with two payload generators:
  \begin{enumerate}[noitemsep,topsep=0pt]
    \item \texttt{genApiKey} which retrieves a key to authenticate with the API; and
    \item \texttt{genCustInfo} to generate the customer information.
  \end{enumerate}
The server is then expected to reply with the response code \texttt{201} indicating that the customer creation was successful, while also including an identification code of type \texttt{String} which is bound with the variable \texttt{custId}. 
Next, the driver chooses between sending \texttt{addCard}, \texttt{addAddr}, \texttt{getCust} or \texttt{deleteCust}, including the data required in the respective payloads by either invoking the specified generators, or use previously-bound variables. 
For instance, in the case where \texttt{addCard} is selected, the payload must include:
  \begin{enumerate*}[noitemsep,topsep=0pt,label=\emph{(\arabic*)}]
    \item the API key stored in the variable \texttt{apiKey},
    \item the specific customer ID stored in the variable \texttt{custId}, and
    \item the card information, by invoking the generator \texttt{getCardInfo}. 
  \end{enumerate*}
  
  The model also specifies the use of a user-provided assertion \texttt{checkCustomer} (line 9) to check the response sent by the \sut for the response \texttt{getCust}. 
The assertion checks whether the retrieved customer information (stored in $c2$) matches with the information provided when the customer was created when \texttt{addCust} was invoked (stored in $c1$). 
If the assertion succeeds, the test driver loops and performs the choice (line 3) once again. 
Otherwise, the test run terminates and the test is marked as failed. 
The recursion repeats until the test driver selects \texttt{delCust}, after which the protocol would terminate successfully.
\qed
\end{example}

\subsection{Semantics of \model}
We give a denotational semantics of  \model by mapping its model
to sets of finite sequences of input/output \emph{events}.
\newcommandx{\psem}[2][2=i,usedefault=@]{\ensuremath{\lsem #1\ifempty{#2}{}{@#2} \rsem}}
Given a output prefix $\pi$, \psem{\pi} denotes the set of values in
the codomain of the generator of the $i$-th parameter of $\pi$ while
for an input prefix $\pi$, \psem \pi is the set of values inhabiting
the type of the $i$-th parameter of $\pi$.
\begin{figure}
  \begin{multline*}
	 \text{for all } i \in I, \pi_i \text{ output prefix, } \vec x_i = x_{i,1},\ldots, x_{i,n_i}
	 \text{ and } \vec v_i = v_{i,1},\ldots, v_{i,n_i}
	 \\
	 \psem{\stSelOp \big\{\pi_i (x_{i,1}:\textsf{G}_{i,1},\ldots,x_{i,n_i}:\textsf{G}_{i,n_i}).{\stS}_i \big\}_{i \in I}}[]
\triangleq\\
    \bigcup_{i \in I} \big\{\stOutOp\pi_i(\vec v_i).r_i
	 \sst
	 \forall 1 \leq j \leq n_i: v_{i,j}\in \psem{\pi_i}[j] \qst
    r_i \in \psem{{\stS}_i[\nicefrac{\vec v_i}{\vec x_i}], \rho}[]\big\}
  \end{multline*}
  \begin{multline*}
	 \text{for all } i \in I, \pi_i \text{ input prefix, } \vec x_i = x_{i,1},\ldots, x_{i,n_i}
	 \text{ and } \vec v_i = v_{i,1},\ldots, v_{i,n_i}
	 \\
	 \psem{\stBraOp \big\{\pi_i (x_{i,1}:\textsf{G}_{i,1},\ldots,x_{i,n_i}:\textsf{G}_{i,n_i}).{\stS}_i \big\}_{i \in I}}[]
\triangleq\\
    \bigcup_{i\in I} \big\{\pi_i(\vec v_{i}).r_i
	 \sst
    \forall 1 \leq j \leq n_i \qst v_{i,j}\in \psem{\pi_i}[j], A_i[\nicefrac{\vec v_{i}}{\vec x_i}]\Downarrow \textsf{tt}, r_i \in \psem{{\stS}_i[\nicefrac{\vec v_i}{\vec x_i}],\rho}[]\big\}
  \end{multline*}
  \[
    \lsem\stRec{X}.\stS,\rho\rsem \triangleq \bigcap \big\{R \ \big|\ \lsem \stS,\rho [X\mapsto R]\subseteq R\rsem\big\} \qquad \lsem X,\rho \rsem \triangleq \rho(X) \qquad \lsem\stEnd,\rho\rsem \triangleq \{\epsilon\}
  \]
  \caption{Semantics for the \model language.}\label{fig:semantics}
  \vspace{-5mm}
\end{figure}
We define 
\begin{align*}
  E = \{\pi(v_1,\ldots,v_n) \sst \pi \text{ is a prefix and }
  \forall 1 \leq i \leq n \qst v_i \in \psem \pi\}
\end{align*}
Let $M$ a \model model and $\rho$ be a map assigning subsets of
$E^\star$ to free recursion variables of $M$.
The semantics of $M$ in $\rho$, is the set $\psem{M,\rho}[]$ defined
according to the equations in \Cref{fig:semantics}.
The semantics is defined by induction on the structure of the \model
models and returns a set of finite traces; the idea is that this set
represents all the executions allowed by a \model specification.
Basically, \psem{M,\rho}[] yields the possible tests that the 
test driver might use.
For example, the internal choice is defined as the set of all possible
output prefixes in the choice with the different variations of the
values that can be generated from the generators.
The case for the external choice is similar, with the only exception
that once the values are replaced in assertions of branches should hold.
The semantics of $\stEnd$ is standard.
The map $\rho$ keeps track of the recursive variables in $\stS$;
initially we assume that $M$ is \emph{closed}, i.e., it has no free
occurrences of recursion variables.
This makes the semantics of recursion standard.

\subsection{Implementation}

We implement the \model language as part of a tool called \tool. 
Implemented in the Scala programming language, takes as input the \model model and the optional preamble, and generates the Scala source code of an executable \emph{test driver} that interacts with the \sut according to the model.
The test driver, in turn, interacts with the REST API exposed by the \sut by using a Scala API which is autogenerated from the provided OpenAPI specification, using OpenAPI Generator.\footnote{\url{https://openapi-generator.tech}}
When the test driver runs, it invokes such Scala API methods to send HTTP requests to the \sut, and to receive and parse its responses; the model determines which requests are sent (and in what order) by the test driver, and which responses are expected.
After completing the test runs, the test driver produces the following output:
\begin{enumerate}[nosep,label=\emph{(\arabic*)}]
\item the level of model coverage achieved by the test runs;
\item a log file with information about every performed test: the random seed used to generate the test, the sequence of requests/responses and their payloads, and the test outcome (pass/fail);
\item an offline representation of failed tests as sequences of \texttt{curl}\footnote{\url{https://curl.se}} commands, which can be executed from a shell to reproduce faults of the \sut without re-executing the test driver. 
\end{enumerate}

\section{Evaluation}\label{sec:evaluation}

In this section, we conduct a comprehensive evaluation of our methodology and its practical application in \tool. This assessment is twofold:
\begin{enumerate}
  \item \textbf{Qualitative Analysis}: We aim to determine if the test-models defined by \model and the test drivers generated by \tool are effective in identifying logic-based faults.
  \item \textbf{Quantitative Analysis}: We assess the effectiveness of our methodology in terms of code coverage, which is a standard metric for evaluating testing tools.
\end{enumerate}
Given the novelty of our approach in user-written, model-based testing for REST-APIs, we establish our baseline comparison with:
\begin{enumerate*}[label=\textit{\roman*)}]
  \item fully-automated REST testing approaches; and
  \item manually crafted REST API tests.
\end{enumerate*}

\subsection{Experiment setup}\label{sec:experiment-setup}
To conduct such evaluation, and obtain the results presented in \Cref{sec:results}, we follow the preparatory steps illustrated in the rest of this section: we first select the artefacts, build their \model model, and we determine an adequate number of test runs per application.

\medskip
\noindent \emph{Artefact selection.}
To conduct such an evaluation, we consider a sample of third-party applications listed in \Cref{tab:apps}, satisfying the following criteria:
\begin{enumerate}
\item they must be open source, to facilitate reproducibility of our results;
  \item they must have OpenAPI specifications, needed by \tool;
  \item they should be non-trivial and thus representative of real-world RESTful applications;
  \item\label{item:criteria:ideal} ideally, they should include manually-written tests of their REST APIs, should be amenable to code coverage measurements, and should have been already used for evaluation in previous literature on testing.
\end{enumerate} 

\begin{table}[t]\centering
  \begin{tabular}[h]{@{}lp{6cm}@{}}
  \toprule
  \textbf{Application}&\textbf{Notes}\\
  \midrule
  \emph{FeaturesService} \cite{features-service}
    &{Used in empirical evaluation of \cite{DBLP:conf/qrs/Arcuri17,DBLP:conf/kbse/StallenbergOP21,DBLP:journals/corr/abs-2205-05325,DBLP:conf/icse/LiuLDLWWJXB22}}\\
  \emph{RESTCountries} \cite{restcountries}&Used in the empirical evaluation of \cite{DBLP:journals/corr/abs-2205-05325}\\
  \emph{GestaoHospital} \cite{gestaohospital}&Used in the empirical evaluation of \cite{DBLP:journals/corr/abs-2205-05325}\\
  \emph{LanguageTool} \cite{languagetool}&Used in the empirical evaluation of \cite{DBLP:journals/corr/abs-2205-05325,DBLP:conf/icse/LiuLDLWWJXB22}\\
  \emph{PetClinic} \cite{petclinic}&{Used in the empirical evaluation of \cite{DBLP:journals/corr/abs-2108-08196}}\\
  \emph{UsersRegistry} \cite{users-regsitry}&{Uses non-trivial API authentication}\\
  \emph{PetStore} \cite{petstore} 
    & Used in the empirical evaluation of \cite{DBLP:conf/icse/LiuLDLWWJXB22} \\
  \emph{SockShop} \cite{sockshop}&{Used in the empirical evaluation of \cite{DBLP:conf/icse/AderaldoMPJ17,DBLP:conf/ecsa/AvritzerFJRSH18}}\\
  \emph{Openverse} \cite{openverse}&Industry app, part of WordPress project\\
  \bottomrule
  \end{tabular}
\caption{Case studies.}\label{tab:apps}
  \vspace*{-5mm}
\end{table}

All applications in \Cref{tab:apps} are open source, provide OpenAPI specifications, and are non-trivial; moreover, all of them (except \emph{UsersRegistry} and \emph{Openverse}) have been used in previous testing literature.
The first 5 applications also satisfy the rest of our ``ideal'' criteria (item~\ref{item:criteria:ideal} above): they include handcrafted test suites for their REST APIs, and their architecture (consisting of a single Java-based executable) allow us to easily collect and analyse code coverage information using standard tools.
Two other applications (\emph{LanguageTool} and \emph{PetStore}) also have similar architecture (allowing us to analyse code coverage), but do not include a test suite for their REST APIs.

\medskip \noindent \emph{Building the \model Models.}
As we are not the authors of the \sut{s} in \Cref{tab:apps}, we were required to infer the usage of the \sut{}s as intended by their developers --- in particular, what sequences of operations (request/responses) are valid, and how some requests may depend on others.
As mentioned earlier, this information is typically only given informally in the application documentation, it may not be up-to-date, and is often omitted. 
Therefore, we often inferred this information by examining the pre-existing handcrafted tests, or by studying existing REST API clients.
We also enriched our test models with application usage scenarios that, although possible, might have been overlooked in the existing handwritten tests. 
This way, our models may leverage \sut functionalities beyond those covered by the existing tests, and potentially reveal new faults in the \sut{}s.

\begin{example}\label{eg:features-service-model}
  \emph{FeaturesService} was the only application with documentation about some of its operations. For instance, the documentation says: 
  \begin{quote}
    \em``The application should allow one to add a constraint such that when a feature requires another feature to be active, the latter feature cannot be deactivated without first deactivating the former.''
  \end{quote}
To test whether \emph{FeaturesService} respects such a description, we formalise the \model model shown in \Cref{lst:features-service} (abridged).
This model specifies that the test driver should first add two features by invoking \lab{addFeature}; their names are created by random generators and bound in variables \stVar{feat1} and \stVar{feat2}. 
Then, the test driver should invoke \lab{addConstraint} specifying that \stVar{feat2} requires \stVar{feat1}. 
Finally, the test driver should attempt to delete \stVar{feat1} via \lab{delFeature}, expecting the \sut to answer with a \lab{400} \quo{bad request} response.
If the \sut was to answer with anything other than a \lab{400}, (such as \lab{200}, indicating that the operation was successful) it would indicate a fault in the logic of the \sut.
  \qed
\end{example}

\begin{figure}
  \small\begin{Verbatim}[commandchars=\\\{\},numbers=left,numbersep=1mm,frame=single,breaklines]
S_featuresService = rec X.(
     +\{ !addFeature(feat1: String(genFeatName)).?C201().
             !addFeature(feat2: String(genFeatName)).?C201().
                 !addConstraint(feat1, feat2).?C201().
                     !delFeature(feat1).?C400().X,
         ...
       \} )
        \end{Verbatim}
\vspace{-3mm}
\caption{\model model for testing \emph{FeaturesServices} (abridged), discussed in \Cref{eg:features-service-model}.}
\label{lst:features-service}
\end{figure}

\medskip \noindent \emph{Determining an adequate number of test runs.}
After writing the \model model of each \sut and generating its
test driver using \tool, we established an adequate number of test runs for each \sut. The optimal number of test runs is the smallest number that maximises
\begin{enumerate*}[label=\emph{(\arabic*)}]
  \item the number of discovered faults, 
  \item the level of code coverage, and
  \item the level of model coverage.
\end{enumerate*}
This number depends on the complexity of the \sut and the \model model in use.
To determine the optimal number, we adopted an incremental approach: gradually increase the number of test runs until the rate of bug discovery and code coverage plateau.
This was determined by analysing the logs generated by \tool for each application and verifying that the model was being fully traversed.

\subsection{Results}\label{sec:results}

We now present our results, in terms of discovered faults
(\Cref{sec:eval:discovered-faults}) and code coverage (\Cref{sec:eval:coverage}).

\subsubsection{Discovered Faults.}\label{sec:eval:discovered-faults}

\Cref{tab:bugs}
reports the number of faults discovered by our test models and the \tool-generated test drivers, categorised by the test oracle that detected the fault.
We found 25 faults across the 9 selected applications and reported some of the most representative faults to the application maintainers;
when we received a reply, in almost all cases the developers acknowledged that the faults were real bugs.\footnote{For \emph{LanguageTool}, we reported 3 cases of \emph{``500 Internal Server Error''}
  uncovered by our test model; such errors are considered faults in REST testing literature,
  but \emph{LanguageTool} produces such errors to signal an unsupported functionality.
}

\Cref{tab:bugs} classifies the faults detected by \tool into logic-based and systematic categories. 
Logic-based faults, embedded in the \sut's logic, are discerned through testing specific operational sequences, while systematic faults, inherent in the \sut's overall function, can be identified by testing the requests in isolation of other requests. 
\tool successfully uncovered 10 logic-based and 15 systematic faults. 
Detecting the logic-based faults is attributed to \model models, which incorporate the \sut's domain knowledge, enabling \tool to navigate complex, interdependent request-response sequences. 
Such intricate sequences are unlikely to be deduced by fully-automated tools (\eg \Cref{eg:features-service-model}), that, as mentioned earlier focus more on systematic faults. 
Furthermore, as \Cref{tab:bugs} demonstrates, \tool effectively detects both fault types.

Some examples of logic-based faults follow: 
\begin{itemize}
  \item In the \emph{PetClinic} model, we included an assertion to verify that a newly created resource can be subsequently retrieved from the \sut. However, \tool discovered a fault: if a resource is updated, subsequent retrieval attempts fail, indicating an issue in the \sut's retrieval operation.
\item Another fault in the \emph{PetClinic} application involved inconsistent identification numbers. The \sut issues an ID number when a resource is created, but the list of resources returned shows different IDs than those originally assigned.
\item The \emph{FeaturesService} application did not adhere to the model shown in \Cref{eg:features-service-model}. It incorrectly permitted the deactivation of \stVar{feat1}, which should have been disallowed, and erroneously returned a successful HTTP status code of \lab{204} instead of indicating failure.
\item In the \emph{SockShop} application, a significant fault was identified: the application fails to retrieve resources immediately after their creation. This fault was detected by first creating the resource and then attempting to retrieve it via the API sequence modelled via \model. 
\end{itemize}

\begin{table}[t]
\begin{tabular}[t]{l@{\quad}c@{\quad}c@{\quad}c@{\quad}|@{\;\;}l|c@{\quad}c@{\;\;}}
  \toprule
  \textbf{Application}&\rotatebox{75}{\parbox{2.3cm}{\bf{Bad status\\code}}}&\rotatebox{75}{\parbox{2.3cm}{\bf{Bad response\\body}}}&\rotatebox{75}{\parbox{2.3cm}{\bf{Assertion\\fail}}}&\rotatebox{75}{\textbf{Total}}&\rotatebox{75}{\parbox{2.3cm}{\bf{Logical\\errors}}}&\rotatebox{75}{\parbox{2.3cm}{\bf{Systematic\\errors}}}\\
  \midrule
  \emph{FeaturesService}&6&0&0&6 (1, 0)&2&4\\
  \emph{RESTCountries}&0&2&0&2 (2, 0)&2&0\\
  \emph{GestaoHospital}&1&0&0&1 (1, 0)&0&1\\
  \emph{LanguageTool}&3&0&0&3 (3, 3)&0&3\\
  \emph{PetClinic}&5&0&2&7 (3, 3)&4&3\\
  \emph{UsersRegistry}&1&0&0&1 (1, 1)&0&1\\
  \emph{PetStore}&1&0&0&1 (1, 0)&0&1\\
  \emph{SockShop}&1&1&0&2 (2, 0)&1&1\\
  \emph{Openverse}&1&1&0&2 (2, 2)&1&1\\
  \bottomrule
  \textbf{Total}&17&5&2&25 (16, 6)&10&15\\
  \end{tabular}
\smallskip
  \caption{Faults discovered by \tool test drivers and oracles, using our \model models.}
  \label{tab:bugs}
  \vspace*{-5mm}
\end{table}

\begin{table*}
  \centering
  \scalebox{1}{\begin{tabular}[t]{@{}l@{\;}|p{1cm}@{\;}p{1cm}@{}|p{1cm}@{\;\;}p{1cm}@{\;\;}p{1.4cm}@{}|p{1.2cm}@{}p{1.2cm}@{}p{1.2cm}@{}}
  \toprule
  \textbf{Application}&\rotatebox{75}{\textbf{Application LOC}}&\rotatebox{75}{\textbf{\#endpoints}}&\rotatebox{75}{\parbox{3cm}{\textbf{\model\\model size (LOC)}}}&\rotatebox{75}{\parbox{3cm}{\textbf{Manual tests\\size (LOC)}}}&\rotatebox{75}{\parbox{3cm}{\textbf{Number of\\\tool test runs}}}&\rotatebox{75}{\parbox{3cm}{\textbf{\tool tests\\line coverage}}}&\rotatebox{75}{\parbox{3cm}{\textbf{Manual tests\\line coverage}}}&\rotatebox{75}{\parbox{3cm}{\textbf{Morest avg.\\line coverage \cite{DBLP:conf/icse/LiuLDLWWJXB22}}}}\\
  \midrule
  \emph{RESTCountries}& 2409 & 22 & 247 & 300 & 100 & \textbf{1722} & 896 &  \\
  \emph{GestaoHospital}& 4427 & 20 & 138 & 463 & 50 & \textbf{2857} & 2532 & \\
  \emph{PetClinic}& 10,416 & 35 & 225 & 1321 & 100 & 3099 & \textbf{3127} &  \\
  \emph{UsersRegistry}& 5452 & 11 & 68 & 246 & 50 & \textbf{2035} & 1906 & \\
  \emph{FeaturesService}& 2026 & 18 & 98 & 377 & 150 & \textbf{1626} & 1576 & 360 \\
  \emph{LanguageTool}& 18,053 & 2 & 75 & & 20 & \textbf{4999} & & 935 \\
  \emph{PetStore}& 3693 & 20 & 111 & & 100 & \textbf{1987} & & 763.20 \\
  \emph{SockShop}& 3392 & 15 & 99 & & 50 & & & \\
  \emph{OpenVerse}& 7117 & 16 & 124 & & 1 & & & \\
  \bottomrule
  \end{tabular}
}\smallskip
  \caption{Size information about the case studies, and comparison between \tool-based testing results (with \model models developed by us, executed for up to 1 minute with default initial randomness seed), handwritten tests (part of the \sut source code), and Morest (average across 5 repetitions, each requiring 8 hours of execution). The coverage is measured using JaCoCo 0.8.8 (\url{https://www.eclemma.org/jacoco}) with standard configuration; JaCoCo is also used in the Morest evaluation \cite{DBLP:conf/icse/LiuLDLWWJXB22}.}\label{tab:comparison}
  \vspace{-5mm}
\end{table*}

\subsubsection{Code Coverage.}\label{sec:eval:coverage}

\Cref{tab:comparison} reports the size of the case studies,\footnote{The \emph{LanguageTool} application has 5 API endpoints but only 2 are testable.}
and compares \tool model-based testing coverage results against handcrafted
REST API test suites (when available as part of the selected case studies), and
against fully-automated REST API testing.

The table shows the sizes (in lines of code) of our \model models against the handwritten REST API test suites.
The \tool model sizes include the preambles, whereas the handwritten test sizes exclude their comments. 
\Cref{tab:comparison} also shows the number of test runs performed using \tool and the number of lines covered by \begin{enumerate*}
  \item \tool;
  \item the applications' handwritten REST API tests; and
  \item the fully-automated tool Morest \cite{DBLP:conf/kbse/LiuLLWWL22, DBLP:conf/icse/LiuLDLWWJXB22}.
\end{enumerate*}
We focus on Morest because a recent study \cite{DBLP:conf/icse/LiuLDLWWJXB22}
shows Morest to be superior to the other fully-automatic REST API testing tools
in literature.
Therefore, we selected Morest as representing the state-of-the-art,
and included some of its case studies (\emph{FeaturesService},
\emph{LanguageTool} and \emph{Petstore}) in our experiments.

In \Cref{tab:comparison} we can observe that \model models are smaller than the handwritten tests provided by the \sut{}s --- and yet, their code coverage is comparable, and higher for \tool in most cases. 
In the case of \emph{RestCountries}, the \model model size is quite close to the handwritten tests size, but the \tool coverage is significantly higher. 
This suggests that a concise \model model can replace (part of) a larger handwritten test suite, allowing testers to ``\emph{concentrate on a (data) model and generation infrastructure instead of hand-crafting individual test}'' \cite{DBLP:conf/icse/DalalJKLLPH99}.

With respect to the fully-automatic testing tool Morest, the benefits of developing a \model model are evident in the significantly higher line coverage achieved by \tool. 
In the three common applications we examined, \tool achieved 3 times as much coverage in \emph{PetStore}, 4 times as much in \emph{FeaturesService}, and 5 times as much in \emph{LanguageTool}. 
Our results were obtained in a few seconds (once the \model model was developed), whereas the Morest coverage is averaged over 5 executions taking 8 hours each.
This suggests that, after the initial investment of time needed to write a \model model, our model-based approach pays off in terms of coverage achieved and time saved over repeated executions.

\subsection{Threats to Validity and Limitations}

The time taken to develop \model models for these case studies varied from one application to another, depending on the complexity of the requests of the applications. 
On average, we estimate that it took us 30 hours to complete each model --- including the time we used to infer the intended usage of the application in question.
We believe that a developer or tester more familiar with the application domain and requirements would write equivalent (or better) test models in less time.

\paragraph{Internal threats to validity.}

The method for determining optimal test runs could lead to imprecisions due to local maxima. Our lack of precise knowledge about expected request/response sequences necessitated designing \model models based on application tests and source code, potentially introducing bias (\Cref{tab:comparison}). Additionally, the comparison does not account for the qualitative distinction between creating concrete test cases and developing more abstract \model models, with the latter possibly being viewed as more challenging.

\paragraph{External threats to validity.}
Our case studies, though diverse, may not universally represent all application types, limiting generalizability (\Cref{sec:experiment-setup}). We aligned our study with REST API testing discourse, selecting artefacts common in literature to minimize bias and ensure relevance. Despite these efforts, the specific selection could limit applicability, though it was necessary for comparability with existing research.

\section{Related Work}\label{sec:related-work}

In contrast to existing methodologies, our \model DSL allows for a more nuanced and comprehensive modelling of REST API behaviour. 
It enables the description of complex sequences and dependencies that are essential for thoroughly testing RESTful services. 
This approach is a significant departure from traditional models, which typically address simpler scenarios or focus on individual API operations in isolation.

We classify studies on testing of REST APIs into two broad categories: \emph{model-based} and \emph{fully automatic}.

\subsubsection{Model-based Testing of REST APIs.}

Our work enriches model-based testing of REST APIs by introducing a sophisticated DSL, \model, for detailed modeling and testing, diverging from existing approaches by supporting complex test sequences and state dependencies within a single model (\Cref{sec:copenapi}). Unlike Chakrabarti \etal \cite{testtherest} and Fertig \etal \cite{DBLP:conf/www/FertigB15}, who focus on isolated sequences or operations, our DSL enables multiple test paths and value assertions, surpassing the limitations of singular test sequence models and isolated API testing. Furthermore, while Seijas \etal \cite{DBLP:conf/erlang/SeijasLT13} and Pinheiro \etal \cite{Pinheiro2013ModelBasedTO} contribute valuable perspectives on property-based and state machine-based modeling, they lack in addressing dynamic data generation and comprehensive evaluation. Francisco \etal's \cite{DBLP:conf/erlang/FranciscoLFC13} constraints-driven approach and Aichernig \etal's \cite{DBLP:journals/sosym/AichernigS19} business rule modeling offer insights into precondition and postcondition specification, yet neither adequately tackles the RESTful API stateful interaction complexity our \model DSL is designed for. Our approach uniquely facilitates capturing intricate dependencies across REST API requests and responses, a critical aspect for thorough testing of sophisticated web services.

\subsubsection{Fully Automatic Testing of REST APIs.}

This category includes a variety of testing approaches \cite{DBLP:conf/edoc/Ed-DouibiIC18, DBLP:conf/kbse/StallenbergOP21, DBLP:journals/tse/SeguraPTC18, DBLP:journals/ese/ZhangMA21} and tools such as RestTestGen \cite{DBLP:conf/icst/ViglianisiDC20}, RESTler \cite{DBLP:conf/icse/AtlidakisGP19}, EvoMaster \cite{DBLP:conf/qrs/Arcuri17}, QuickREST \cite{DBLP:conf/icst/KarlssonCS20}, RestCT \cite{DBLP:conf/icse/WuXNN22} and Morest \cite{DBLP:conf/icse/LiuLDLWWJXB22, DBLP:conf/kbse/LiuLLWWL22}.
Starting from an API specification like OpenAPI \cite{openAPI}, these tools automatically generate tests primarily to check the correctness of individual API responses. 
They mostly rely on random data generators or generators automatically derived from the specifications, which might not fully capture the intricacies of the API's behaviour.
In contrast, in our work, the sequencing of valid requests and responses is explicitly specified in the model, enabling complex interactions with the \sut. 
This is complemented by our use of non-random, tester-assigned data generators in the \model model, which allows for a more targeted and effective testing approach. 
This method provides a stark contrast to AGORA's invariant detection-based approach \cite{DBLP:conf/issta/AlonsoSR23} and the approach in \cite{DBLP:journals/tsc/Martin-LopezSMR22} focussing on the specification of inter-parameter dependencies. 
While these methodologies offer robust solutions in their respective areas, they do not offer the same level of specificity in interaction sequencing and tailored data generation as our \model model-driven approach does.

\section{Conclusion}\label{sec:concl}

We introduced \tool, a tool implementing a model-based testing approach for RESTful applications using \model DSL and OpenAPI specifications, which generates executable test drivers for model-based testing of web services. Our evaluation against third-party open source applications and comparison with existing automated REST API testing tools and manual test suites demonstrates the efficacy of \tool. It achieved comparable or better code coverage and fault detection with smaller, more manageable models, highlighting the potential for reduced effort in test creation and maintenance. Future work includes extending \tool to support other API specifications like GraphQL and gRPC, and enhancing \model's expressiveness by incorporating additional test oracles and timing constraints.

\subsubsection*{Acknowledgements.}
We thank the anonymous reviewers for their comments. This
work was partially supported by: the Horizon Europe grant 101093006 (TaRDIS);
the BehAPI project funded by the EU H2020 RISE under the Marie Skłodowska-Curie action (No: 778233); the PRIN 2022 PNRR project DeLiCE (F53D23009130001);
the MUR project PON REACT EU; the PRO3 MUR project Software Quality; ``the
MUR dipartimento di eccellenza;'' the PNRR MUR project VITALITY (ECS00000041);
Spoke 2 ASTRA - Advanced Space Technologies and Research Alliance.

\bibliographystyle{splncs04}
\bibliography{refs.bib}

\end{document}